\documentclass[11pt]{amsart}
\usepackage{enumitem}
\usepackage{setspace}
\usepackage{natbib}
\usepackage{geometry}                
\usepackage{graphicx}
\usepackage{amssymb}
\usepackage{amsmath}
\usepackage{epstopdf}
\usepackage[linesnumbered,vlined,algoruled]{algorithm2e}
\usepackage{url}
\usepackage{mathtools}

\title{The Size of a $t$-Digest}
\author{Ted Dunning}
\address{Ted Dunning \\ MapR Technologies, Inc \\ San Jose, CA}
\email{ted.dunning@gmail.com}
\date{}                                           

\begin{document}
\begin{abstract}
A $t$-digest is a compact data structure that allows estimates of quantiles which increased accuracy near $q = 0$ or $q=1$. This is done by clustering samples from $\mathbb R$ subject to a constraint that the number of points associated with any particular centroid is constrained so that the so-called $k$-size of the centroid is always $\le 1$. The $k$-size is defined using a scale function that maps quantile $q$ to index $k$. This paper provides bounds on the sizes of $t$-digests created using any of four known scale functions.
\end{abstract}
\maketitle
\section{Introduction}
We examine four cases for the $t$-digest \citep{t-digest-arxiv} scale function in order to produce estimates of the number of centroids and the maximum size of these centroids. The functions include
\[
\begin{aligned}
k_0(q) &= \frac \delta 2 q \\
k_1(q) &= \frac \delta {2\pi}  \sin^{-1}(2q-1)   \\
k_2(q) &= \frac \delta {Z(n)} \log {\frac q {1-q}} \\
k_3(q) &= \frac \delta {Z(n)}\begin{cases}
\quad \log 2q & \text{if  } q \le 1/2 \\
- \log 2(1-q) & \text{if  } q > 1/2
\end{cases}
\end{aligned}
\]
Where $Z$ is a normalization value that may depend on the number of data points seen so far and is intended to bound the size of the $t$-digest, at least in a practical sense.
\section{General considerations}
\begin{figure}[htbp] 
   \centering
   \includegraphics[width=2.3in]{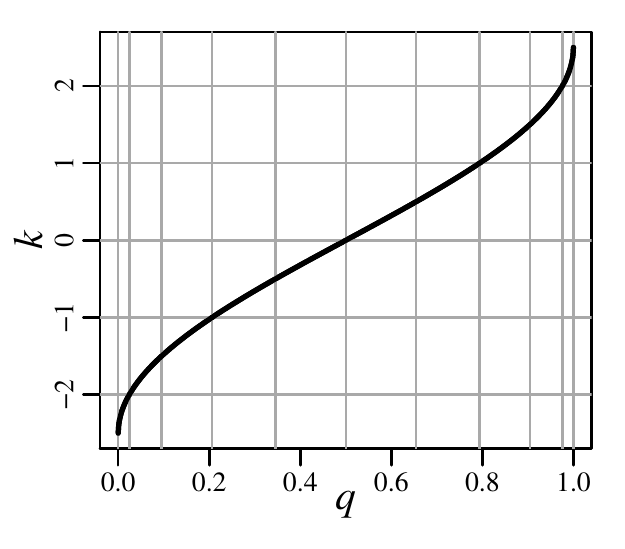} 
   \caption{The scaling function translates the quantile $q$ to the scaling factor $k$ in order to give variable size steps in $q$. This limits the number of samples that are included in any particular centroid, particularly near $q=0$ or $q=1$ which in turn allows more accurate estimation of the distribution near these limits. }
   \label{fig:k-q-plot-full}
\end{figure}

Figure \ref{fig:k-q-plot-full} shows how a scaling function converts quantiles (the $q$ axis) into $k$ values. In this figure, values of $k$ are marked  and how unit steps in $k$ cause variable sized steps in $q$. This limit is the fundamental idea behind $t$-digests together with the idea of representing the distribution of interest by a set of centroids, each of which summarizes a localized subset of the samples. Making some subsets smaller means that the error due to summarizing the distribution is reduced near the extreme values of $q$.

These boundaries in $q$ are, however, not precisely realizable in practice since the centroids are made up of an integral number of samples. As such, centroids are limited in size to be smaller than the idealized size, except in the case of single samples. Importantly, centroids with only a single point may have $k$-size larger than $1$. Such centroids do not, however, contribute to excess error since we know the exact position of such lone centroids.

\begin{figure}[htbp] 
   \centering
   \includegraphics[width=3.5in]{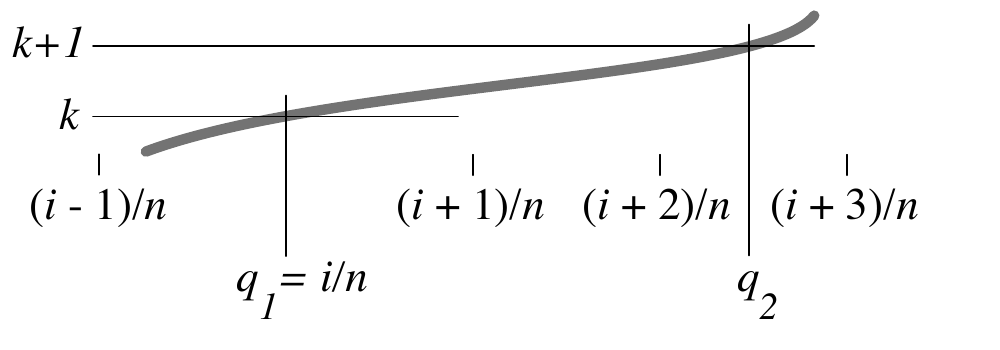} 
   \caption{Looking in detail at a magnified part of the scaling function, taking unit steps in $k$ gives idealized boundaries for centroids, but the weights of centroids are  constrained to integral values. }
   \label{fig:k-q-limits}
\end{figure}

Figure \ref{fig:k-q-limits} illustrates how this works out in practice. Taking the case of one particular centroid $\mathcal C$, then if we assume that the samples that contribute to all of the centroids with smaller mean values are less than the samples that define $\mathcal C$ and similarly that all of the samples for centroids with larger mean values are to the right of all of the samples in $\mathcal C$, we can make some statements about how many samples can be in $\mathcal C$. Depending on how the $t$-digest was constructed, this assumption may not be quite true, but it will be close to true.
For the purpose of this illustration, there are $i$ samples in centroids with smaller means than $\mathcal C$ and we mark this as point $q_1$. Point $q_1$ maps to the value $k$ and $k+1$ maps back to $q_2$ which is the limit on the size of the centroid. Here it is clear $\mathcal C$ can contain only two samples without extending past $q_2$ and thus being too large to be a valid part of a $t$-digest.

\begin{figure}[htbp] 
   \centering
   \includegraphics[width=3.5in]{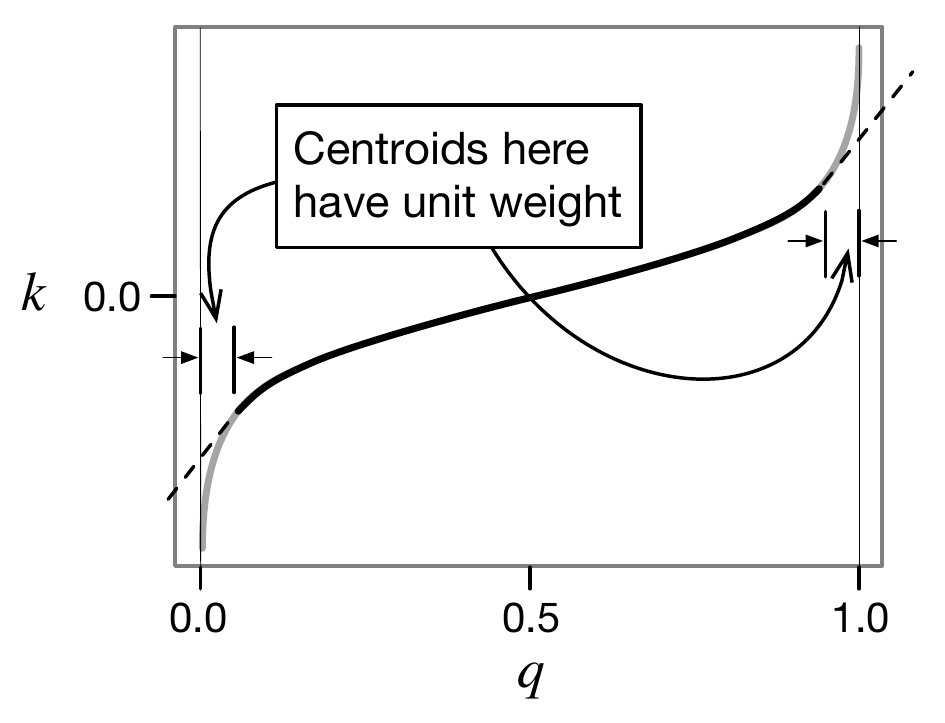} 
   \caption{When the slope becomes too steep, centroids have unit weight. The effect is that the scaling function (black and gray thick line) is effectively modified (dashed lines) so that the slope is at most $n$, the total number of samples.  }
   \label{fig:k-q-slope}
\end{figure}

As can be seen from these figures, if the scaling curve is too steep, $q_2-q_1 < 1/n$ so $\mathcal C$ will have only a single sample. This observation turns out to be very important when calculating how many samples a centroid will use. Figure \ref{fig:k-q-slope} shows the effect of this. All practically important scaling functions become very steep near the boundaries so there will be some number of unit weight centroids at the ends of any $t$-digest. 

Exactly how many such unit centroids there are likely to be is a very important factor in characterizing the $t$-digest and can be computed by finding points where the scaling function exceeds a critical slope. Once a step from $q=i/n$ to $q'=(i+1)/n$ results in an increase in $k$ greater than one, no centroid can have more than one sample. The critical slope is thus $n$, the total number of samples.

In the next sections, these ideas will be used to determine specific limits on $t$-digest size for each of the important scaling functions listed at the beginning of this paper.

\section{Size bound for $k_0$}
The slope of $k_0$ is constant and whenever that slope is more than $n$, all centroids will have unit weight. This happens whenever $n \le \delta/2 $ so for this case the number of centroids is just $n$. 

In a mid-range where $\delta/2 < n < \delta$, the $k$-size of unit centroids is less than $1$, but the $k$-size of a centroid with weight more than $1$ is larger than $1$ so no centroids can combine without violating the $t$-digest size limit. Thus the number of centroids is $n$ in this range as well.

If we take $  n = \mu \delta /2$  where $\mu \ge 2$ all centroids will have an average weight of at least $ \mu/2 $ but none will have a weight larger than $ \mu $ and thus $\delta/2 \le n \le \delta$.

In any of these three cases, for scaling function $k_0$ the number of centroids in a  $t$-digest will be in the range $\left[\min( n,\delta/2), \min(n,\delta)\right]$.

Based on the logic above, the upper bound for the weight of any centroid for $k_0$ is $
w \le 2 n / \delta$. 
\section{Size bound for $k_1$}
We know that $k_1(1) -  k_1(0) = \delta/2$. Since no centroid can have $k$-weight more than $1$ and the average $k$-weight cannot be less than $1/2$, we know that the number of centroids is in $\left[ \delta/2 , \delta \right]$ if $n > \delta$. 

The slope of $k_1(q)$ is $\delta / ( 2\pi \sqrt{q(1-q)} \, )$. The minimum slope is 
$\delta / \pi$ at $q=1/2$ so if $n \le \delta/\pi$, no centroid will have a weight of more than one.

For $\delta/\pi < n \le \delta$ the slope at $1/2$ will be such that some centroids can have a weight more than one.
\subsection{Maximum centroid weight}

The maximum weight of centroids can be computed by considering a centroid centered in terms of $k$ around $k(q) = \kappa$. 

This centroid can span a range of $q$ of at most
\[
\Delta q = k^{-1}(\kappa+1/2) - k^{-1}(\kappa-1/2) \\
\]
But we know that
\[
k^{-1}(\kappa) = \frac 1 2 \left(   \sin \frac {2\pi \kappa} \delta + 1  \right)
\]
So using multiple angle identities
\[
\begin{aligned}
\Delta q &= \frac 1 2 \left(
  { \sin \frac {2\pi (\kappa+1/2)} \delta  }  -   { \sin \frac {2\pi (\kappa-1/2)} \delta  } 
  \right) \\
&= \frac 1 2 \Biggl( 
\left( \sin \frac {2 \pi} \delta \kappa \cos \frac \pi \delta  + \cos \frac {2 \pi} \delta \kappa \sin \frac \pi \delta  \right) -
\left( \sin \frac {2 \pi} \delta \kappa \cos \frac  \pi  \delta  - \cos \frac {2 \pi} \delta \kappa \sin \frac  \pi  \delta  \right) 
\Biggr) \\
&=  
\sin \frac \pi \delta    \cos \frac {2 \pi} \delta \kappa =   2 \left( \sin \frac \pi \delta \right)    \sqrt{q(1-q)}\\
\end{aligned}
\]
The maximum weight of this centroid is  $w_{\max} = n \Delta q$.

Note, however, the value of $q$ cannot not be known precisely from the information we have in a $t$-digest for a centroid. If, indeed, we did know lower and upper bounds $q_{\min}$ and $q_{\max}$ for the points associated with a centroid in a $k_1$ digest, we would be able to use whichever value of $q$ that is furthest from $1/2$ to get a conservative limit $w_{\max}$.
\[
q_{\text {worst}} = \begin{cases}
 q_{\min} & \text{if } | q_{\min} - 1/2 | > | q_{\max} - 1/2 | \\
 q_{\max} & \text {otherwise}
\end{cases} \\
\]
Unfortunately, it is not possible to know the actual values of $q_{min}$ and $q_{max}$ in most practical situations such as when a $t$-digest has been created by merging digests. For a particular centroid $\mathcal C$, if we take $\mathcal W_{\text {left}}$ to be the sum of the weights of centroids with means smaller than $\mathcal C$ and $\mathcal W$ to be the weight of $\mathcal C$ then these estimated values are useful
\[
\begin{aligned}
q_{\min} &\approx {\mathcal W_{left}} / n \\
q_{\max} &\approx (\mathcal W + \mathcal W_{left}) / n  
\end{aligned}
\]
From that 
\[
w_{\max} \le 2n \left( \sin \frac \pi \delta\right) \sqrt{q_{\text {worst}} (1-q_{\text {worst}})}
\]
Note that in any case, we have no solid guarantees about the distribution of the samples for a particular centroid so that this size limit remains somewhat heuristic in nature.
\section{Size bounds for $k_2$}
Unlike $k_1$, both $k_2$ and $k_3$ have unbounded tails at $q=0$ and $q=1$ so we require a different argument to compute the size of a $t$-digest for $k_2$ and $k_3$. Because the smallest weight a centroid can have is $1$ corresponding to a change in $q$ of $\Delta q = 1/n$, the effective slope of the scaling function is limited to a value of $n$ and thus the range from minimum to maximum $k$  is finite. For suitably chosen  $Z(n)$, the size is not only finite, but bounded for any value of $n$.

Wherever $k_2'(q) > n$, a centroid cannot have a weight greater than $1$. In particular, if the smallest value of the slope is this large, all centroids must have unit weight. By definition,
\[
\begin{aligned}
k_2(q) &= \frac \delta {Z(n)} \log {\frac q {1-q}} \\
k_2'(q) &= {\frac  {\delta/Z(n)} {q (1-q) }  } \\
\end{aligned}
\]
The slope is least at $q=1/2$ so if $ n < 4  \delta / Z$ all centroids have unit weight.

On the other hand, even if $n$ is considerably larger than this limit, some centroids near $q=0$ or $q=1$ will still have unit weight because the slope of $k(q)$ increases without bound and thus will ultimately exceed any value of $n$ for $q$ sufficiently near $0$ or $1$. We can approximate the region $[q_1, q_2]$ where centroids can have weights larger than $1$ by examining where the slope of $k(q)$ crosses the critical value of $n$.
\[
\begin{aligned}
{\frac  {\delta/Z(n)} {q_i (1-q_i) }  } &= n \\
q_i^2 - q_i + \frac {\delta/Z(n)} n &= 0 \\
q_i &= \frac 1 2 \left( { 1 \pm \sqrt { 1 - 4 \frac {\delta/Z(n)} n } } \, \right)
\end{aligned}
\]
The discriminant here will always be positive for the values $n$ we are examining so we have two solutions symmetric around $1/2$. Moreover, if $n \gg 4\delta / Z$
\[
\begin{aligned}
\sqrt { 1 - 4 \frac {\delta/Z(n)} n } &\approx 1 - 2 \frac {\delta / Z(n)} n \\
q_1 &\approx {   \frac {\delta/Z(n)} n  } \,  \\
q_2 &= 1-q_1
\end{aligned}
\]
This means that there will be roughly $\delta/Z(n)$ unit centroids at each extreme with the remaining centroids having a weight greater than one. We can compute an approximate bound on the number of larger centroids
\[
\begin{aligned}
m_{w>1} &\le 2\left( k(q_2) -k(q_1) \right) \\ 
&\le  \frac {2\delta} {Z(n)} \log {\frac {q_2 (1-q_1)} {(1-q_2) q_1}} = \frac {2\delta} {Z(n)} \log { {q_2^2} / { q_1^2}} \\
&\approx  \frac {4 \delta} {Z(n)} \left(\log  n/ \delta + \log Z(n)\right)
\end{aligned}
\]
Adding back in the unit clusters
\[
m \le  \frac {4 \delta} {Z(n)} \left(\log  n/ \delta + \log Z(n) + 1/2\right)
\]

If we pick $Z(n) = 1$, then the total number of centroids grows with increasing $n$ and is roughly
\[
m \le 4\delta  \log n/\delta + 1/2
\]  
On the other hand, if we pick $Z(n)$ so that it grows with $\log n$ then $\log Z(n)$  will grow extremely slowly. Specifically, choosing $Z(n) = 4 \log n/\delta + 24$ means that 
\[
\frac {\log n / \delta + \log Z(n)+1/2} {Z(n)}  < 1/4
\]
for all values of $n/\delta < 9.1 \times 10^{23}$ which more than covers the range of practical utility.

This inequality means that the total number of centroids will be bounded 
\[
\begin{aligned}
m&\le   \frac {4 \delta} {Z(n)} \left(  \log  n/ \delta + \log Z(n) + 1/2\right) \\
&\le  \delta
\end{aligned}
\]
\subsection{Maximum centroid weight}

The maximum weight for a centroid centered at $k(q) = \kappa$ can be estimated using a Taylor expansion
\[
\begin{aligned}
w_{\max} &\le n \Delta q \\
\Delta q &= k^{-1}(\kappa + 1/2) - k^{-1}(\kappa-1/2) \\
&\approx \left . \frac {dq} {dk} \right\rvert_{\kappa} = \left[ 1 \Bigg \slash {\frac {dk} {dq}} \right]_q\\
&\approx \frac {Z(n)} \delta q(1-q) \\
\end{aligned}
\]
As before, in practice, we cannot know the precise value of $q = k^{-1}(\kappa)$ so we typically use $q_{\text {worst}}$ to get a conservative limit.
\section{Size bounds for $k_3$}
The scale function $k_3$ is
\[
k_3(q) = \frac \delta {Z(n)}\begin{cases}
\quad \log 2q & \text{if  } q \le 1/2 \\
- \log 2(1-q) & \text{if  } q > 1/2
\end{cases}
\]
By symmetry, we can consider just the first branch in our calculations. In this range, the slope is
\[
k_3'(q) = \frac \delta { q Z(n)}  
\]
All centroids with $q \le \frac \delta {2 n Z(n)}$ have unit weight. When $n \le \frac \delta {Z(n)}$, this region extends to $q = 1/2$ forcing all centroids to have unit weight. For larger $n$, we define $q_1$ as before (and $q_2$ by symmetry, of course)
\[
\begin{aligned}
q_1 &= \frac \delta { n Z(n)} \\
q_2 &= 1-q_1
\end{aligned}
\]
The total number of centroids is thus
\[
\begin{aligned}
m  &\le  n q_1 + n (1-q_2) + 2(k(q_2) - k(q_1)  ) \\
  &\le 2\left(  n q_1 + (k(q_2) - k(q_1)  ) \right)\\
&= \frac { 2\delta} { Z(n)} \left(1  -2\log 2 q_1\right)\\
&= \frac {4\delta} { Z(n)} \left(1/2- \log 2 + \log  \frac  { n }\delta + \log Z(n)\right)
\end{aligned}
\]
For $Z(n)=1$, we have unbounded logarithmic growth in the number of centroids. Alternatively, if we choose $Z(n) = 4\log n/\delta + 21$  and take $n/\delta < 10^{20}$ as a reasonable restriction on the area of applicability, we get a bound of $m \le \delta$.
\subsection{Maximum centroid weight}
We can follow the same method as with $k_2$ to find the maximum weight for any centroid.

The maximum weight for a centroid centered at $k(q) = \kappa$ can be estimated using a Taylor expansion
\[
\begin{aligned}
w_{\max} &\le n \Delta q \\
\Delta q &= k^{-1}(\kappa + 1/2) - k^{-1}(\kappa-1/2) \\
&\approx \left . \frac {dq} {dk} \right\rvert_{\kappa} = \left[ 1 \Bigg \slash {\frac {dk} {dq}} \right]_q\\
&\approx \frac {Z(n)} \delta \min (q, (1-q)) \\
\end{aligned}
\]
As before, $q$ is not known precisely in practice so we use $q_{\text {worst}}$ to get a conservative limit.
\[
w_{\max}\le \frac {nZ(n)} \delta \min (q_{\text{worst}}, (1-q_{\text{worst}})) 
\]
\bibliographystyle{chicago}
\bibliography{refs}{}

\end{document}